\documentclass[runningheads]{llncs}

\usepackage{algorithmicx}
\usepackage[noend]{algpseudocode}
\usepackage{algorithm}

\algnewcommand{\algorithmicswitch}{\textbf{switch}}
\algdef{SE}[SWITCH]{Switch}{EndSwitch}[1]{\algorithmicswitch\ #1\ \algorithmicdo}{\algorithmicend\ \algorithmicswitch}%
\algtext*{EndSwitch}%

\algnewcommand{\algorithmiccase}{\textbf{case}}
\algdef{SE}[CASE]{Case}{EndCase}[1]{\algorithmiccase\ #1}{\algorithmicend\ \algorithmiccase}%
\algtext*{EndCase}%

\algnewcommand{\algorithmicon}{\textbf{on}}
\algdef{SE}[ON]{On}{EndOn}[1]{\algorithmicon\ #1\ \algorithmicdo}{\algorithmicend\ \algorithmicon}%
\algtext*{EndOn}%

\algrenewcommand{\algorithmicdo}{}
\algrenewcommand{\algorithmicthen}{}

\algnewcommand{\algorithmicgoto}{\textbf{goto}}%
\algnewcommand{\Goto}[1]{\algorithmicgoto~\ref{#1}}%

\algnewcommand{\algorithmicbreak}{\textbf{break}}%
\algnewcommand{\Break}[0]{\algorithmicbreak}%

\algnewcommand{\algorithmicwaiton}{\textbf{wait on}}%
\algnewcommand{\WaitOn}[1]{\algorithmicwaiton~{#1}}%

\PassOptionsToPackage{hyphens}{url}
\usepackage{url}
\usepackage{hyperref}
\hypersetup{breaklinks=true}

\usepackage[utf8]{inputenc}
\usepackage[T1]{fontenc}

\usepackage{amssymb,amsfonts,amsmath,amsthm}
\usepackage{graphicx}
\usepackage{xcolor}

\usepackage[caption=false,font=footnotesize]{subfig}
\usepackage{float}

\usepackage{IEEEtrantools}
\allowdisplaybreaks

\usepackage[shortlabels,inline]{enumitem}

\usepackage{xstring}

\usepackage{datetime2}

\usepackage{siunitx}

\usepackage{varwidth}

\usepackage{tikz}
\usetikzlibrary{calc}
\usetikzlibrary{arrows}
\usetikzlibrary{arrows.meta}
\usetikzlibrary{patterns}
\usetikzlibrary{positioning}
\usetikzlibrary{decorations.pathreplacing}
\usetikzlibrary{shapes.misc}
\usetikzlibrary{spy}

\pgfdeclarelayer{bg1}
\pgfdeclarelayer{bg2}
\pgfsetlayers{bg1,bg2,main}

\usepackage{pgfplots}
\pgfplotsset{compat=1.14}
\usepgfplotslibrary{fillbetween}

\usepackage{xspace}
\usepackage{ifthen}

\newcommand{\LOGda}[2]{%
    \ifthenelse{\equal{#1}{}}{%
        \ensuremath{\mathsf{LOG}_{\mathrm{da}}^{#2}}%
    }{%
        \ensuremath{\mathsf{LOG}_{\mathrm{da},#1}^{#2}}%
    }%
}
\newcommand{\LOGbft}[2]{%
    \ifthenelse{\equal{#1}{}}{%
        \ensuremath{\mathsf{LOG}_{\mathrm{bft}}^{#2}}%
    }{%
        \ensuremath{\mathsf{LOG}_{\mathrm{bft},#1}^{#2}}%
    }%
}

\newcommand{\ie}[0]{\emph{i.e.}\xspace}
\newcommand{\eg}[0]{\emph{e.g.}\xspace}
\newcommand{\cf}[0]{\emph{cf.}\xspace}

\newcommand{\wolog}[0]{w.l.o.g.\xspace}

\theoremstyle{plain}
\newtheorem*{theorem*}{Theorem}

\definecolor{myParula01Blue}{RGB}{0,114,189}
\definecolor{myParula02Orange}{RGB}{217,83,25}
\definecolor{myParula03Yellow}{RGB}{237,177,32}
\definecolor{myParula04Purple}{RGB}{126,47,142}
\definecolor{myParula05Green}{RGB}{119,172,48}
\definecolor{myParula06LightBlue}{RGB}{77,190,238}
\definecolor{myParula07Red}{RGB}{162,20,47}

\tikzset{myparula11/.style={color=myParula01Blue,solid,mark=+,mark options={solid}}}
\tikzset{myparula12/.style={color=myParula01Blue,densely dashed,mark=x,mark options={solid}}}
\tikzset{myparula13/.style={color=myParula01Blue,densely dotted,mark=o,mark options={solid}}}
\tikzset{myparula14/.style={color=myParula01Blue,dashdotted,mark=triangle,mark options={solid}}}
\tikzset{myparula15/.style={color=myParula01Blue,dashdotdotted,mark=square,mark options={solid}}}

\tikzset{myparula21/.style={color=myParula02Orange,solid,mark=+,mark options={solid}}}
\tikzset{myparula22/.style={color=myParula02Orange,densely dashed,mark=x,mark options={solid}}}
\tikzset{myparula23/.style={color=myParula02Orange,densely dotted,mark=o,mark options={solid}}}
\tikzset{myparula24/.style={color=myParula02Orange,dashdotted,mark=triangle,mark options={solid}}}
\tikzset{myparula25/.style={color=myParula02Orange,dashdotdotted,mark=square,mark options={solid}}}

\tikzset{myparula31/.style={color=myParula03Yellow,solid,mark=+,mark options={solid}}}
\tikzset{myparula32/.style={color=myParula03Yellow,densely dashed,mark=x,mark options={solid}}}
\tikzset{myparula33/.style={color=myParula03Yellow,densely dotted,mark=o,mark options={solid}}}
\tikzset{myparula34/.style={color=myParula03Yellow,dashdotted,mark=triangle,mark options={solid}}}
\tikzset{myparula35/.style={color=myParula03Yellow,dashdotdotted,mark=square,mark options={solid}}}

\tikzset{myparula41/.style={color=myParula04Purple,solid,mark=+,mark options={solid}}}
\tikzset{myparula42/.style={color=myParula04Purple,densely dashed,mark=x,mark options={solid}}}
\tikzset{myparula43/.style={color=myParula04Purple,densely dotted,mark=o,mark options={solid}}}
\tikzset{myparula44/.style={color=myParula04Purple,dashdotted,mark=triangle,mark options={solid}}}
\tikzset{myparula45/.style={color=myParula04Purple,dashdotdotted,mark=square,mark options={solid}}}

\tikzset{myparula51/.style={color=myParula05Green,solid,mark=+,mark options={solid}}}
\tikzset{myparula52/.style={color=myParula05Green,densely dashed,mark=x,mark options={solid}}}
\tikzset{myparula53/.style={color=myParula05Green,densely dotted,mark=o,mark options={solid}}}
\tikzset{myparula54/.style={color=myParula05Green,dashdotted,mark=triangle,mark options={solid}}}
\tikzset{myparula55/.style={color=myParula05Green,dashdotdotted,mark=square,mark options={solid}}}

\tikzset{myparula61/.style={color=myParula06LightBlue,solid,mark=+,mark options={solid}}}
\tikzset{myparula62/.style={color=myParula06LightBlue,densely dashed,mark=x,mark options={solid}}}
\tikzset{myparula63/.style={color=myParula06LightBlue,densely dotted,mark=o,mark options={solid}}}
\tikzset{myparula64/.style={color=myParula06LightBlue,dashdotted,mark=triangle,mark options={solid}}}
\tikzset{myparula65/.style={color=myParula06LightBlue,dashdotdotted,mark=square,mark options={solid}}}

\tikzset{myparula71/.style={color=myParula07Red,solid,mark=+,mark options={solid}}}
\tikzset{myparula72/.style={color=myParula07Red,densely dashed,mark=x,mark options={solid}}}
\tikzset{myparula73/.style={color=myParula07Red,densely dotted,mark=o,mark options={solid}}}
\tikzset{myparula74/.style={color=myParula07Red,dashdotted,mark=triangle,mark options={solid}}}
\tikzset{myparula75/.style={color=myParula07Red,dashdotdotted,mark=square,mark options={solid}}}

\pgfplotsset{
    mysimpleplot/.style = {
        every axis plot/.prefix style={thick},
        width=1.0\linewidth,
        height=0.75\linewidth,
        title style={font=\footnotesize,align=center},
        legend cell align=left,
        legend style={font=\footnotesize},
        legend columns=3,
        legend style={
            at={(0.5,1)},
            yshift=0.3em,
            anchor=south,
            draw=none,
            /tikz/every even column/.append style={
                column sep=0.3em
            },
            cells={
                align=left
            }
        },
        grid=both,
        minor tick num=3,
        major grid style={solid,draw=gray!50},
        minor grid style={densely dotted,draw=gray!50},
        label style={font=\footnotesize,align=center},
        tick label style={font=\footnotesize},
    },
}

\begin{document}
\title{Three Attacks on Proof-of-Stake Ethereum}
\author{Caspar Schwarz-Schilling\inst{1} \and Joachim Neu\inst{2} \and Barnabé Monnot\inst{1} \and Aditya Asgaonkar\inst{1} \and Ertem Nusret Tas\inst{2} \and David Tse\inst{2}}
\authorrunning{C. Schwarz-Schilling, J. Neu, B. Monnot, A. Asgaonkar, E. N. Tas, D. Tse}
\institute{Ethereum Foundation\\\email{\{caspar.schwarz-schilling,barnabe.monnot,aditya.asgaonkar\}@ethereum.org} \and Stanford University\\\email{\{jneu,nusret,dntse\}@stanford.edu}}
\maketitle              %
\begin{abstract}
Recently, two attacks were presented against Proof-of-Stake (PoS) Ethereum: one where short-range reorganizations of the underlying consensus chain are used to increase individual validators' profits and delay consensus decisions, and one where adversarial network delay is leveraged to stall consensus decisions indefinitely. We provide refined variants of these attacks, considerably relaxing the requirements on adversarial stake and network timing, and thus rendering the attacks more severe. Combining techniques from both refined attacks, we obtain a third attack which allows an adversary with vanishingly small fraction of stake and no control over network message propagation (assuming instead probabilistic message propagation) to cause even long-range consensus chain reorganizations. Honest-but-rational or ideologically motivated validators could use this attack to increase their profits or stall the protocol, threatening incentive alignment and security of PoS Ethereum. The attack can also lead to destabilization of consensus from congestion in vote processing.
\end{abstract}

\section{Introduction}
\label{sec:introduction}

The Proof-of-Stake (PoS) Ethereum consensus protocol \cite{eth2-spec-beaconchain,eth2-spec-forkchoice,eth2-spec-validator}
is constructed by applying the finality gadget
Casper FFG \cite{casper} on top of the fork choice rule LMD GHOST,
a flavor of the Greedy Heaviest-Observed Sub-Tree (GHOST) \cite{ghost} rule
which considers only each participant's most recent vote
(Latest Message Driven, LMD).
Participants with stake that allows them to vote as part of the protocol
are called \emph{validators}.
A slightly simplified and analytically more tractable variant of PoS Ethereum
is given by the Gasper protocol \cite{gasper}.

Recent works \cite{lowcostreorgs,ebbandflow,ethresearch-balancing-attack} have presented two attacks on Gasper and PoS Ethereum.
The first attack \cite{lowcostreorgs}
uses
short-range reorganizations (\emph{reorgs}) of the blockchain
stipulating consensus 
to
delay finality of consensus decisions.
Such short-range reorgs also allow
validators to increase their earnings
from participating in the protocol
(\eg, from Maximal Extractable Value, MEV \cite{flashboys}).
As a result, honest-but-rational validators will deviate from the protocol,
threatening the assumptions underlying the security arguments for it.
In the second attack \cite{ebbandflow,ethresearch-balancing-attack},
the adversary exploits 
adversarial network delay
and strategic voting by a vanishing fraction of adversarial validators
to stall the protocol indefinitely.

\paragraph{Our Contributions}

In this paper we present enhanced variants of the above two attacks
\cite{lowcostreorgs,ebbandflow}.
First, we reduce the number of validators necessary to launch a short-range reorg.
An adversary who could perform a reorg of $k$ blocks (\emph{$k$-reorg}) using the old strategy \cite{lowcostreorgs}
is now able to perform a $(k+1)$-reorg using our new strategy.
Second, we considerably relax the network assumption under which the adversary can stall
PoS Ethereum using techniques similar to \cite{ebbandflow,ethresearch-balancing-attack}:
we show that the adversary does not need to exert control over message propagation delays,
but that merely stationary \emph{probabilistic} network delay,
as is commonly assumed to model networks under normal operation,
together with a still vanishingly small (albeit slightly larger than before)
fraction of adversarial validators
suffices
for the adversary to be able to
effectively stall the protocol.
We then combine techniques from both refined attacks
to devise a long-range reorg attack
which requires only an extremely small number of adversarial validators
and no adversarial (but only probabilistic) network delay.

This third attack is particularly severe for PoS Ethereum for three reasons:
\begin{enumerate*}
    \item Honest-but-rational validators might adopt the strategy
        as they can use it to increase their
        payouts from MEV and transaction fees.
        The resulting protocol deviations destabilize consensus
        on both the fork choice and the finality gadget level
        because the blockchain does not grow steadily anymore.
        
    \item Reorgs lead to uncertainty and delay in block confirmation,
        impacting user experience and quality of service, and undermining
        users' trust in the protocol.
        
    \item Reorgs can reduce the throughput of the consensus layer
        to the point where not enough votes can be processed timely,
        reducing resilience against adversarial validators and
        jeopardizing proper functioning of PoS Ethereum.

\end{enumerate*}

\paragraph{Related Work}

In both selfish mining \cite{selfishmining}
and our attacks the adversary withholds blocks to displace honest blocks from the chain. Unlike selfish mining however, our attacks do not lead to an increased block production reward.
Undercutting attacks \cite{whaletx} showcase how consensus instability can arise from reorgs incentivized by large variance in block rewards.
In fact, this concern will be aggravated by diminishing block rewards in Bitcoin in the future
\cite{bitcoininstability}.
Time-bandit attacks \cite{flashboys} point out that MEV earned in past blocks can incentivize and subsidize reorgs and other attacks in the future,
\eg, for renting hash power or bribing validators.

\paragraph{Outline}

PoS Ethereum and its network model are reviewed in Section~\ref{sec:preliminaries}.
Sections~\ref{sec:refined-reorg-attack} and~\ref{sec:refined-liveness_attack} each first
introduce a recent
attack and then describe our refined variant thereof.
Combining techniques from our refined attacks,
we devise a long-range reorg attack in Section~\ref{sec:doubly-refined-reorg-attack}.
We
discuss
in Section~\ref{sec:discussion}
the impact of the presented long-range reorg attack
on various aspects of PoS Ethereum.

\section{Proof-of-Stake Ethereum: The Gasper Protocol}
\label{sec:preliminaries}
We provide a concise summary of the PoS Ethereum/Gasper protocol and the network environment it is designed for.
The exposition is slightly idealized and streamlined for ease of comprehension.
For all details, 
refer to
the paper \cite{gasper} of Gasper
and
the PoS Ethereum beacon chain protocol specifications \cite{eth2-spec-forkchoice,eth2-spec-validator,eth2-spec-beaconchain}.

\subsection{Model}
\label{sec:prelims-model}

We assume a static pool of $N$ protocol participants (called \emph{validators} or \emph{nodes}), each with unit stake. This corresponds to consensus in a \emph{permissioned} setting. Network communication among validators is synchronous, \ie, network delay is under adversarial control, up to a known delay upper bound $\Delta$. Clocks across nodes are synchronized. This amounts to a \emph{synchronous network} \cite{model-sync}. There is an external shared source of randomness which can be used by the protocol to sample a group (of predetermined size) of validators in a uniform manner without replacement. Validators follow the protocol as prescribed, except for a fraction $\beta$ which are under adversarial control and can deviate from the protocol in arbitrary and coordinated fashion (\emph{Byzantine faults}).

In its basic version, the state machine replication (SMR) formulation of consensus
asks for a protocol
that can be run among the $N$ protocol participants
to obtain a linear ordering of \emph{transactions} input by the environment
to participants, into a shared \emph{ledger}
(\ie, to implement an ordering service)
with the following security properties:
\begin{itemize}
    \item \emph{Liveness:} If some honest validator becomes aware of a transaction,
        then not too long thereafter
        that transaction will have entered the ledger as output by any honest validator
        (\ie, `good things do happen', `transactions enter the ledger').
    \item \emph{Safety:} The ledgers output by different honest validators at different points
        in time are consistent. In other words, it does not happen that a transaction,
        which has once
        entered the ledger in some honest validator's view at some time, disappears later
        (\ie, `bad things do not happen', `if a transaction enters the ledger, then it will not leave it').
\end{itemize}

Given an SMR protocol, we seek to understand for which adversarial fractions $\beta$ the ledger output by that protocol is both safe and live (and hence \emph{secure}).

\subsection{Protocol}
\label{sec:prelims-protocol}

Being a composite with the LMD GHOST fork choice rule as the basis
and Casper FFG as a finality gadget on top,
PoS Ethereum consensus proceeds roughly in two stages and on two time scales.

First, on the smaller time scale where LMD GHOST operates, time proceeds in synchronized slots of duration $2\Delta$.
For each slot, one \emph{block proposer}
and a \emph{committee} of $W$ validators is drawn uniformly at random
from the $N$ validators.
The following LMD GHOST rule is used to determine a canonical block (and its prefix of blocks as a canonical chain) in a node's view in slot $t$: ``Starting at the highest block $b_0$ `justified' by Casper FFG (see below), sum for each child block $b$ the number of unique (\ie, one per slot and slot's committee member, breaking ties adversarially) valid (\ie, only from earlier than the current slot, and no voting on future blocks) votes for that block and its descendants; count for every validator only its most recently cast vote (LMD). Pick the child block $b^*$ with highest weight (GHOST) (breaking ties adversarially). Recurse ($b_0 \leftarrow b^*$), until reaching a leaf block. Output that leaf block.''
At the beginning of each slot, the slot's proposer determines a block using LMD GHOST and extends it with a new proposal.
Half way into each slot (\ie, $\Delta$ time after the proposal and after the beginning of the slot), the slot's committee members determine a block using LMD GHOST in their view and vote for it (votes are also called \emph{attestations}).
(At the same time they also cast a Casper FFG vote, as described later.)
An exact confirmation rule of LMD GHOST/Gasper is not specified.

Second, on the larger time scale where Casper FFG operates, time proceeds in epochs comprised of $32$ slots.
On a high level, Casper FFG is a two-phase traditional propose-and-vote-style Byzantine fault tolerant (BFT) consensus protocol (cast as a blockchain protocol into the chained framework, like Chained HotStuff \cite{yin2018hotstuff}), except there is no leader in charge of assembling proposals. Instead, the proposals are supposed to be generated consistently across honest nodes by the LMD GHOST fork choice layer.
Casper FFG proceeds as follows: Blocks first become justified if a super-majority ($2N/3$) votes `for them', and subsequently become finalized, roughly when a super-majority votes `from them' for a subsequent block. The genesis block is justified and finalized by definition.
The blocks among which validators cast their votes during an epoch are the so-called epoch boundary blocks, which are those blocks that are leaf blocks after truncating the block tree to only those blocks that came from the previous epoch.
Validators vote for the highest epoch boundary block that is consistent with the highest justified block they have observed, which in turn extends the latest finalized block they have observed.
Due to the super-majority required to advance a proposal, as well as the two-phase confirmation (called \emph{finalization}), Casper FFG remains safe even under temporary network partition.
The confirmation rule on the Casper FFG level is to output the latest finalized block and its prefix.

\section{A Refined Reorg Attack}
\label{sec:refined-reorg-attack}
\subsection{Motivation}

Previous work \cite{lowcostreorgs} described a malicious, low-cost reorg attack. In particular, the attack leverages strategic timing of broadcasting blocks and attestations, as opposed to honestly releasing them when supposed to. In a nutshell, in the strategy of \cite{lowcostreorgs}, an adversarial block proposer in slot $n$ keeps its proposal hidden. The honest block proposer in slot $n+1$ will then propose a competing block. The adversary can now use its committee members' votes from both slots $n$ and $n+1$ to vote for the withheld block of slot $n$ in an attempt to outnumber honest votes on the proposal of slot $n+1$. As a result, blocks proposed by honest validators may end up orphaned, \ie, they are displaced out of the chain chosen by LMD GHOST. In \cite{lowcostreorgs} this reorg strategy is part of a bigger scheme to delay consensus.

We show how the attack of \cite{lowcostreorgs} can be modified such that the number of adversary validators required is significantly reduced, from a set of size linear in the total number of validators to a constant-size set -- indeed for a one-block reorg as little as one adversarial validator is sufficient. Note that similar to \cite{lowcostreorgs} the adversarial strategy does not involve any slashable behavior and is therefore relatively cheap. In Section~\ref{sec:doubly-refined-reorg-attack}, we further improve upon this refined reorg attack, combining strategies from both this section and Section~\ref{sec:refined-liveness_attack}.

\subsection{Refined Reorg Strategy}

\begin{figure}[t]
    \centering
    \begin{tikzpicture}[
        _block/.style = {
            draw,
            minimum width=3.5em,
            minimum height=3.5em,
        },
        honestblock/.style = {
            _block,
            fill=green!50!black!10,
        },
        adversarialblock/.style = {
            _block,
            fill=red!15,
        },
        _vote/.style = {
            circle,
            draw,
            inner sep=0,
            minimum width=0.8em,
            minimum height=0.8em,
            font=\tiny,
        },
        honestvote/.style = {
            _vote,
            fill=green!50!black!10,
        },
        adversarialvote/.style = {
            _vote,
            fill=red!15,
        },
        hidden/.style = {
            dashed,
        },
        sloteven/.style = {
            fill=black!5,
            draw=none,
        },
        arrowvotes/.style = {
            -Latex,
            shorten <=3pt,
            shorten >=3pt,
            densely dotted,
        },
        groupvotes/.style = {
            decorate,
            decoration={
                brace,
                amplitude=3pt,
                raise=3pt,
            },
        },
        groupvotestip/.style = {
            midway,
            inner sep=0,
            yshift=7pt,
        },
        x=2.5cm,
        y=0.75cm,
    ]
        \footnotesize
    
        \draw [sloteven] (-0.5,2.9) rectangle (0.5,-3.5);
        \node at (0,2.4) {\textsc{Slot $n$}};
        \node at (1,2.4) {\textsc{Slot $n+1$}};
        \draw [sloteven] (1.5,2.9) rectangle (2.5,-3.5);
        \node at (2,2.4) {\textsc{Slot $n+2$}};
        \node at (3,2.4) {\textsc{Slot $n+3$}};
        
        \draw [dotted] (-1,-3) -- (3.5,-3);
        \node [fill=white] at (-0.9,-2.95) {\textsc{Votes}};
    
        \node [honestblock] (b1) at (0,0) {$n$};
        \node [adversarialblock,hidden] (b2) at (1,-0.9) {$n+1$};
        \node [honestblock] (b3) at (2,0.9) {$n+2$};
        \node [honestblock] (b4) at (3,-0.9) {$n+3$};
        
        \draw [-Latex,dashed] (b1) -- (-0.75,0);
        \draw [-Latex,bend left=5] (b2) to (b1);
        \draw [-Latex,bend right=5] (b3) to (b1);
        \draw [-Latex] (b4) -- (b2);
        
        \node [honestvote] (v11) at ($(0,-3)+(-0.35,0)$) {};
        \node [honestvote] (v12) at ($(0,-3)+(-0.21,0)$) {};
        \node [honestvote] (v13) at ($(0,-3)+(-0.07,0)$) {};
        \node [honestvote] (v14) at ($(0,-3)+(+0.07,0)$) {};
        \node [honestvote] (v15) at ($(0,-3)+(+0.21,0)$) {};
        \node [honestvote] (v16) at ($(0,-3)+(+0.35,0)$) {};
        
        \draw [groupvotes]
            (v11.north west) -- (v16.north east) node [groupvotestip] (q11) {};
        \draw [arrowvotes] (q11) -- (b1);
        
        \node [honestvote] (v21) at ($(1,-3)+(-0.35,0)$) {};
        \node [honestvote] (v22) at ($(1,-3)+(-0.21,0)$) {};
        \node [honestvote] (v23) at ($(1,-3)+(-0.07,0)$) {};
        \node [honestvote] (v24) at ($(1,-3)+(+0.07,0)$) {};
        \node [honestvote] (v25) at ($(1,-3)+(+0.21,0)$) {};
        \node [adversarialvote,hidden] (v26) at ($(1,-3)+(+0.35,0)$) {};
        
        \draw [groupvotes]
            (v21.north west) -- (v25.north east) node [groupvotestip] (q21) {};
        \draw [arrowvotes] (q21) -- (b1);
        \draw [groupvotes]
            (v26.north west) -- (v26.north east) node [groupvotestip] (q22) {};
        \draw [arrowvotes] (q22) -- (b2);
        
        \node [honestvote] (v31) at ($(2,-3)+(-0.35,0)$) {};
        \node [honestvote] (v32) at ($(2,-3)+(-0.21,0)$) {};
        \node [honestvote] (v33) at ($(2,-3)+(-0.07,0)$) {};
        \node [honestvote] (v34) at ($(2,-3)+(+0.07,0)$) {};
        \node [honestvote] (v35) at ($(2,-3)+(+0.21,0)$) {};
        \node [honestvote] (v36) at ($(2,-3)+(+0.35,0)$) {};
        
        \draw [groupvotes]
            (v31.north west) -- (v36.north east) node [groupvotestip] (q31) {};
        \draw [arrowvotes] (q31) -- (b2);
        
        \node [honestvote] (v41) at ($(3,-3)+(-0.35,0)$) {};
        \node [honestvote] (v42) at ($(3,-3)+(-0.21,0)$) {};
        \node [honestvote] (v43) at ($(3,-3)+(-0.07,0)$) {};
        \node [honestvote] (v44) at ($(3,-3)+(+0.07,0)$) {};
        \node [honestvote] (v45) at ($(3,-3)+(+0.21,0)$) {};
        \node [honestvote] (v46) at ($(3,-3)+(+0.35,0)$) {};
        
        \draw [groupvotes]
            (v41.north west) -- (v46.north east) node [groupvotestip] (q41) {};
        \draw [arrowvotes] (q41) -- (b4);

    \end{tikzpicture}
    \caption[]{Example of a one-block reorg attack using the refined strategy: In slot $n+1$ the adversary privately creates block $n+1$ on block $n$ and attests to it. Honest validators of slot $n+1$ do not see any block and thus attest to block $n$ as head of the chain. In the next slot, an honest proposer publishes block $n+2$ building on block $n$, which is the current head in their view. Simultaneously, the adversary finally publishes block $n+1$ and the attestation voting for block $n+1$. All honest validators of slot $n+2$ attest to block $n+1$ as head of the chain, because it has more weight than block $n+2$. In the next slot block $n+3$ is proposed building on block $n+1$. Block $n+2$ is reorged out.%
    }
    \label{fig:exante-reorg}
\end{figure}
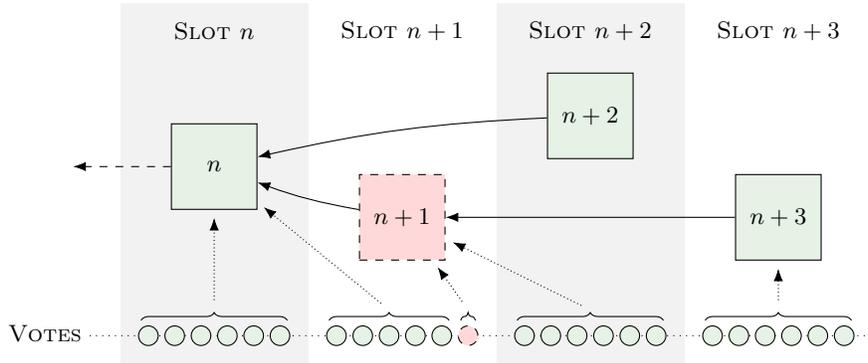

Consider Figure \ref{fig:exante-reorg}, which shows the adversary being the proposer of slot $n+1$ as well as controlling a committee member in slot $n+1$. We describe the adversarial strategy to perform a $1$-reorg: 
\begin{enumerate}
    \item At the beginning of slot $n+1$ the adversary privately creates block $n+1$ on block $n$ and privately attests to it. Honest validators do not see block $n+1$ and so they attest to the previous head of the chain, block $n$. %
    \item At the beginning of the next slot, an honest validator proposes block $n+2$. Assuming zero network latency for now, the adversary finally publishes the private block and attestation from slot $n+1$ at the same time as block $n+2$ is released. Honest validators now see both block $n+1$ (and its one attestation) as well as block $n+2$. These blocks are conflicting because they share the same parent, block $n$. Another result of sharing the same parent is that block $n+1$ inherits all the weight of block $n$, in particular the honest attestations from slot $n+1$ voting for block $n$ also count in favor of it.
    \item Hence, in slot $n+2$ all honest validators vote for block $n+1$ as head of the chain, because it has more weight due to the single adversarial attestation from slot $n+1$. 
    \item Finally, at the beginning of slot $n+3$, an honest validator proposes block $n+3$ pointing to block $n+1$ as its parent. This effectively orphans block $n+2$ and brings the reorg attack to its conclusion.
\end{enumerate}

The above strategy shows that a block proposer which controls a single committee member of the same slot can successfully perform a $1$-reorg. Naturally, the logic of this strategy can be extended to reorg attacks of arbitrary length $k$. Let the number of honest validators in any given committee be $W_{\mathrm{honest}} \approx (1-\beta) W \leq W$. Then, for a successful reorg attack of length $k>1$, the proposing adversary needs to control $W_{\mathrm{honest}} (k-1) + 1$ validators, since it offsets honest committee members' votes in the first $(k-1)$ slots and uses the above refined attack strategy in the last slot.

The refined reorg attacked described here improves on the strategy proposed in \cite{lowcostreorgs} by removing the need for the adversary to compete with the committee of slot $n+k+1$. While the improvement for long-range reorg attacks may not be as significant, short reorg attacks are considerably more feasible using the above refined strategy. In particular, $1$-reorg attacks are effectively always possible for large enough parties. With currently $230{,}000$ active validators\footnote{\url{https://beaconcha.in/validators}. Accessed: 2021-10-09} and $32$ slots per epoch, an adversary controlling $200$ validators (which amounts to $0.09\%$ of total stake) has a $99.8\%$ chance of being selected block proposer at least once per any given day, and once selected as block proposer in a particular slot controls at least one committee member validator in that slot with probability $99.8\%$. So with more than $99.6\%$ probability, an adversary with $0.09\%$ of total stake is in a position to execute a $1$-reorg for any given day.

We will now relax the assumption of zero network latency. PoS Ethereum's fork choice rule only considers attestations that are at least one slot old \cite{eth2-spec-forkchoice} (so votes from slot $n+2$ do not count in the fork choice for slot $n+2$). Further, a committee member is supposed to attest if ``(a) the validator has received a valid block from the expected block proposer for the assigned slot or (b) one-third of the slot has transpired [...] – whichever comes first''\footnote{Regarding attestation timing, PoS Ethereum practice slightly deviates from Gasper} \cite{eth2-spec-validator}. After block $n+2$ is broadcasted to the network, honest validators immediately attest to it upon reception (unless by that time they see another chain as leading in fork choice). Thus, the adversary must ensure that a majority of validators of slot $n+2$ see block $n+1$ and the adversary's attestation voting for block $n+1$ (from slot $n+1$) before they see block $n+2$, but after block $n+2$ was proposed (to ensure it extends block $n$). This proves to be a non-trivial but practically feasible issue.

Suppose the adversary controls a number of nodes at different `locations' in the topology of the peer-to-peer gossip network \cite{eth2-spec-p2p} (these nodes might still be physically collocated). This is possible without greater difficulty because the gossip network has no defenses against such Sybil attacks. Then, some adversarial node will likely receive the new proposal block $n+2$ relatively early on in its dissemination process. The adversary can then release the private block and attestation in a coordinated fashion from all the different locations in the peer-to-peer topology where the adversary controls nodes. Due to the superior number of sources of the adversarial block and attestation it is likely that these arrive earlier than the proposal block $n+2$ at enough (a majority of) honest nodes to ultimately orphan block $n+2$.

\section{A Refined Liveness Attack}
\label{sec:refined-liveness_attack}
\subsection{Motivation}
\label{sec:appendix-attack-gasper-motivation}

Earlier works
\cite{ethresearch-bouncing-attack,ethresearch-bouncing-attack-analysis,ethresearch-bouncing-attack-prevention,ethresearch-balancing-attack,ebbandflow}
have described
balancing-type attacks against variants of the GHOST fork choice rule used in PoS Ethereum
as modelled in the Gasper protocol \cite{gasper}.
In particular, the attack described in \cite{ebbandflow,ethresearch-balancing-attack}
uses adversarial network delay to show that PoS Ethereum is not secure
in traditional (partially) synchronous
networks.
While adversarial network delay (up to some delay bound)
is a
widely employed assumption in the consensus literature,
there is disagreement whether it is appropriate for
Internet-scale open-participation consensus.
As a result, past attacks are often seen as impractical and have not been mitigated: %
``Note that this attack does depend on networking assumptions that are highly contrived in practice (the attacker having fine-grained control over latencies of individual validators), [...]'' \cite{mitigationlmdghostbalancingattacks}

We show how the attack of \cite{ethresearch-balancing-attack,ebbandflow}
can be modified and implemented
\cite{ethresearch-balancing-attack2}
so that
an adversary controlling 15\% of stake
can stall PoS Ethereum
\emph{without requiring adversarial network delay}.
(For ever larger numbers of validators, ever smaller
fractions of adversarial stake suffice.)
To this end, we show through experiments that aggregate properties
of
many individually random message propagation processes
(\eg, `within time $T$ this transmission is received by fraction $x$ of nodes')
in real-world Internet-scale
peer-to-peer gossip networks \cite{eth2-spec-p2p,gossipsub}
are sufficiently predictable
to give the adversary
the required control over how many validators see which
adversarial messages when.
None of the adversarial actions are slashable protocol violations.

\subsection{High-Level Idea}
\label{sec:appendix-attack-gasper-high-level}

Recall
that the \emph{balancing attack} \cite{ethresearch-balancing-attack,ebbandflow} consists of two steps:
First, adversarial block proposers
initiate two competing chains -- call them Left and Right.
Then, a handful of adversarial votes per slot,
released under carefully chosen
circumstances,
suffice to steer honest validators' votes so as to keep the system
in a tie between the two chains and consequently stall consensus.

\begin{figure}[t]
    \centering
    \begin{tikzpicture}[xscale=0.9,yscale=0.75]
        \footnotesize
        \draw [-Latex] (-4.5,0) -- (8,0) node[below] {Time};
        \draw [Latex-] (-3,0) -- (-3,4.75) node[below right] {Adversary sends sway vote for Right from slot $\leq i-1$};
        \draw [Latex-] (-1,0) -- (-1,4) node[below right] {$A$ receives sway vote};
        \draw [Latex-] (0,0) -- (0,3.25) node[below right] {Honest validators scheduled to vote for slot $i$};
        \draw [Latex-] (0.15,0) -- (0.15,2.5) node[below right] {$A$ votes Right (due to sway vote)};
        \draw [Latex-] (0.3,0) -- (0.3,1.75) node[below right] {$B$ votes Left (due to tie-break)};
        \draw [dotted] (0,-1.5) -- (0,5);
        \draw [Latex-] (2,0) -- (2,1) node[below right] {$B$ receives sway vote};
        \draw [Latex-Latex] (-3,-0.25) -- (0,-0.25) node[pos=0.7,below] {$T_{\mathrm{delay}}$};
        \draw [Latex-Latex] (-3,-1) -- (-1,-1) node[midway,above] {$T_{A}$};
        \draw [Latex-Latex] (-3,-1.2) -- (2,-1.2) node[midway,below] {$T_{B}$};
    \end{tikzpicture}
    \caption{Assuming a tie between two chains Left and Right, with tie-break favoring Left.
        The adversary releases a sway vote for Right from a slot $< i$ at time $T_{\mathrm{delay}}$
        before the point in time at which honest validators vote in slot $i$ according to the protocol.
        The parameter $T_{\mathrm{delay}}$ is chosen such that roughly half of honest validators (such as $A$)
        receive the sway vote \emph{before} they submit their vote (and hence vote Right,
        as Right \emph{now} has more votes \emph{in their view}), and the other half of honest validators (such as $B$) receive
        the sway vote \emph{after} they submit their vote for (and hence vote Left,
        as the tie-break \emph{still} favors Left \emph{in their view}).}
    \label{fig:timing-attack}
\end{figure}
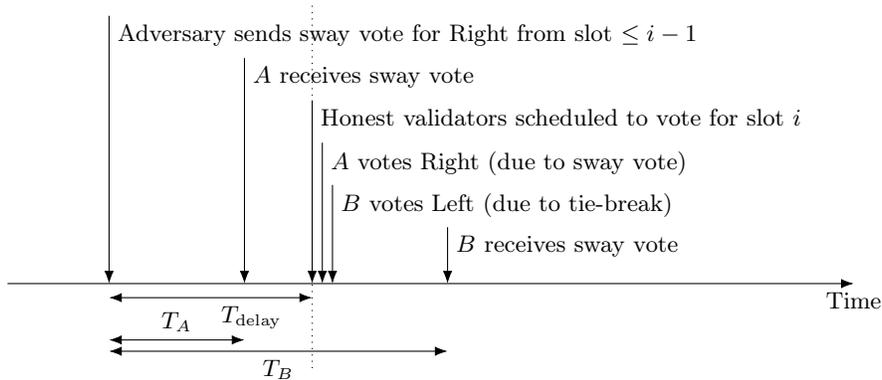

Assume, \wolog, that when viewing Left and Right
with equal number of votes,
the protocol's tie-break favors Left over Right.
If the adversary manages to deliver a withheld adversarial vote for Right from
an earlier slot
to roughly one half of honest validators for the current slot $i$, before validators submit
their votes for slot $i$, while the other half
does not receive
said vote before casting their votes,
then roughly half of honest validators
(those who have received the sway vote `in time')
see Right as leading and will vote for it in slot $i$,
while the other half
(those who see the sway vote `late' and hence at the time of voting see a tie
which they break in favor of Left)
will vote for Left in slot $i$ (see Figure~\ref{fig:timing-attack}).

Idealizing the above
as voting according to a coin flip for each validator,
roughly $W_{\mathrm{honest}}/2$ of $W_{\mathrm{honest}}$ honest validators per slot would vote
Left and Right, respectively,
with a gap of $O(\sqrt{W_{\mathrm{honest}}})$
(\cf variance of a binomially distributed random variable).
So, $O(1/\sqrt{W_{\mathrm{honest}}})$ adversarial fraction of stake would suffice
to rebalance the vote to a tie and keep the system in limbo.
In Section~\ref{sec:appendix-attack-gasper-experiments} we provide evidence from real-world propagation
delay measurements in a replica of Ethereum 2's gossip network \cite{eth2-spec-p2p} to support the hypothesis
that the adversary can indeed reliably determine the time $T_{\mathrm{delay}}$ it takes for approximately half of nodes
to receive a message broadcast by the adversary.

\subsection{Detailed Description}
\label{sec:appendix-attack-gasper-details}

First we describe the attack for a given $T_{\mathrm{delay}}$,
then we describe how
to obtain $T_{\mathrm{delay}}$.
Our simulation\footnote{Source code: \url{https://github.com/tse-group/gasper-gossip-attack}}
using the gossip network propagation model obtained
in Section~\ref{sec:appendix-attack-gasper-experiments}
provides further details.

First, the adversary waits for an opportune epoch to launch the attack. An epoch is opportune
if the block proposers in slot $0$ and $1$ are adversarial (this can be strengthened). Due to the random committee selection in PoS Ethereum,
this happens with probability $\beta^2$ for any given epoch, so that the adversary needs to wait
on average $1/\beta^2$ epochs until it can launch the attack.
In the following, assume epoch $0$ is opportune.
The adversarial proposers of slots $0$ and $1$ propose
conflicting new chains `Left' and `Right', respectively.
Note that this is not a slashable protocol violation.
Both withhold their proposals so that none of slot $0$ or $1$ honest validators
vote for either block.
The adversary releases the blocks
after slot $1$. We assume \wolog
that the tie between Left and Right
(recall that no vote has been cast for either so far)
is broken in favor of Left.

Time $T_{\mathrm{delay}}$ before honest validators in slot $2$
vote, the adversary
releases a vote for Right from an adversarial committee member of slot $1$ (so called \emph{sway vote},
see Figure~\ref{fig:timing-attack}).
If $T_{\mathrm{delay}}$ is tuned well to the network propagation behavior
\emph{at large}, then roughly one half of honest committee members of slot $2$ see the sway vote before
they cast their vote, and thus view Right as leading
(due to the sway vote) and will vote for it;
and the other half
see the sway vote only after they cast their
vote, and thus view Left as leading (due to the tie-break) at the time of voting
and will vote for it.
Once the adversary has observed the outcome of the vote, which now should be a split
up to an $O(\sqrt{W_{\mathrm{honest}}})$ gap, the adversary uses its slot $2$ committee members
(which stipulates the adversarial fraction $O(1/\sqrt{W_{\mathrm{honest}}})$ required
for this attack)
as well as slot $0$ and $1$ committee members to rebalance the vote to a tie.
As the tie is restored, the adversary can use the same strategy in the following slot,
and so forth.

Note that the adversary can observe the outcome of a vote
and learns
how many honest committee members saw Left and Right leading, respectively.
The adversary can use this information to improve its estimate
of $T_{\mathrm{delay}}$.
We show in Section~\ref{sec:appendix-attack-gasper-experiments} that the optimal $T_{\mathrm{delay}}$ can be reliably
localized using grid search.

\subsection{Experimental Evaluation}
\label{sec:appendix-attack-gasper-experiments}

\begin{figure}[t]
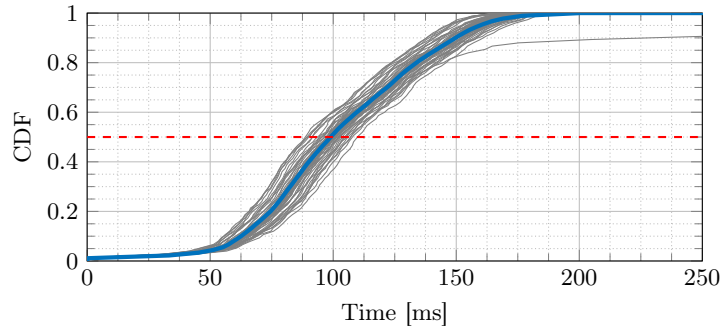

    \centering
    \begin{tikzpicture}
        \small
        \begin{axis}[
            mysimpleplot,
            xlabel={Time [ms]},
            ylabel={CDF},
            xmin=0, xmax=250,
            ymin=0, ymax=1,
            height=0.4\linewidth,
            width=0.8\linewidth,
        ]
        
            \input{figures/gossip-propagation-experiments/node-0-delay-cdfs-50-samples}
            \input{figures/gossip-propagation-experiments/node-0-delay-cdfs-mean}

            \addplot [red,dashed,no marks] coordinates {
                (0, 0.5) (1000,0.5)
            };

        \end{axis}
    \end{tikzpicture}%
    \vspace{-0.3em}
    \caption{Fraction of participants in the peer-to-peer gossip network
        who have received a message broadcast
        by node $0$ at time $0$ by the given time
        ($50$ sample messages in gray, mean over all samples in blue).
        Median (dashed red) at $\approx100\,\mathrm{ms}$.}
    \label{fig:cdfs-0}
\end{figure}

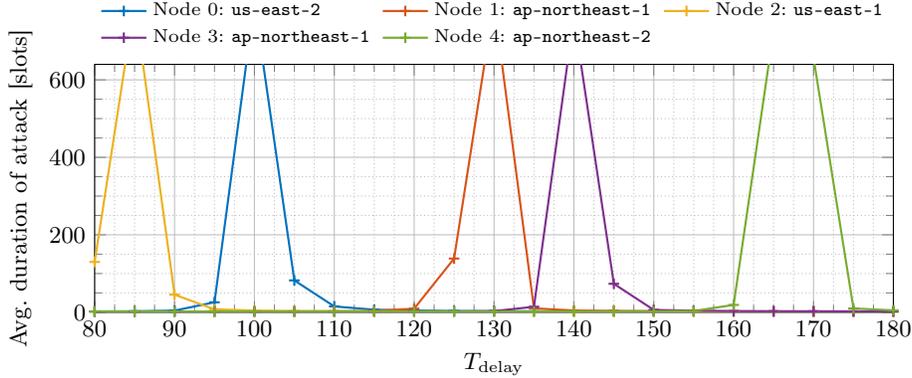
\begin{figure}[t]
    \centering
    \begin{tikzpicture}
        \small
        \begin{axis}[
            mysimpleplot,
            xlabel={$T_{\mathrm{delay}}$},
            ylabel={Avg. duration of attack [slots]},
            legend columns=3,
            xmin=80, xmax=180,
            ymin=0, ymax=640,
            height=0.4\linewidth,
            width=\linewidth,
        ]

            \addplot [myparula11] coordinates {
                (80.0, 2.2) (85.0, 2.6) (90.0, 4.3) (95.0, 25.8) (100.0, 799.0) (105.0, 82.3) (110.0, 15.1) (115.0, 6.6) (120.0, 4.6) (125.0, 3.2) (130.0, 2.8) (135.0, 2.6) (140.0, 2.5) (145.0, 2.4) (150.0, 2.1) (155.0, 2.1) (160.0, 2.1) (165.0, 2.1) (170.0, 2.0) (175.00000000000003, 2.0) (180.0, 2.0)
            };
            \addlegendentry{\scriptsize Node $0$: \texttt{us-east-2}}

            \addplot [myparula21] coordinates {
                (80.0, 2.0) (85.0, 2.0) (90.0, 2.0) (95.0, 2.0) (100.0, 2.1) (105.0, 2.2) (110.0, 2.5) (115.0, 3.5) (120.0, 9.3) (125.0, 138.7) (130.0, 799.0) (135.0, 10.2) (140.0, 4.5) (145.0, 3.5) (150.0, 3.0) (155.0, 2.6) (160.0, 2.4) (165.0, 2.2) (170.0, 2.1) (175.00000000000003, 2.1) (180.0, 2.1)
            };
            \addlegendentry{\scriptsize Node $1$: \texttt{ap-northeast-1}}

            \addplot [myparula31] coordinates {
                (80.0, 129.9) (85.0, 799.0) (90.0, 45.4) (95.0, 7.7) (100.0, 4.5) (105.0, 3.2) (110.0, 2.9) (115.0, 2.6) (120.0, 2.4) (125.0, 2.1) (130.0, 2.0) (135.0, 2.0) (140.0, 2.0) (145.0, 2.0) (150.0, 2.0) (155.0, 2.0) (160.0, 2.0) (165.0, 2.0) (170.0, 2.0) (175.00000000000003, 2.0) (180.0, 2.0)
            };
            \addlegendentry{\scriptsize Node $2$: \texttt{us-east-1}}

            \addplot [myparula41] coordinates {
                (80.0, 2.0) (85.0, 2.0) (90.0, 2.0) (95.0, 2.0) (100.0, 2.0) (105.0, 2.0) (110.0, 2.0) (115.0, 2.0) (120.0, 2.2) (125.0, 2.4) (130.0, 3.0) (135.0, 14.2) (140.0, 763.2) (145.0, 73.5) (150.0, 6.1) (155.0, 4.1) (160.0, 3.0) (165.0, 2.7) (170.0, 2.6) (175.00000000000003, 2.4) (180.0, 2.2)
            };
            \addlegendentry{\scriptsize Node $3$: \texttt{ap-northeast-1}}

            \addplot [myparula51] coordinates {
                (80.0, 2.0) (85.0, 2.0) (90.0, 2.0) (95.0, 2.0) (100.0, 2.0) (105.0, 2.0) (110.0, 2.0) (115.0, 2.0) (120.0, 2.0) (125.0, 2.0) (130.0, 2.0) (135.0, 2.0) (140.0, 2.0) (145.0, 2.1) (150.0, 2.4) (155.0, 3.3) (160.0, 19.0) (165.0, 726.6) (170.0, 656.9) (175.00000000000003, 9.9) (180.0, 4.9)
            };
            \addlegendentry{\scriptsize Node $4$: \texttt{ap-northeast-2}}

        \end{axis}
    \end{tikzpicture}%
    \vspace{-0.3em}
    \caption{Using the propagation delay measurements to model network propagation,
        we simulated our attack
        for fixed $\beta=0.15$, varying $T_{\mathrm{delay}}$,
        and five different positions of the adversary in the network,
        and plot the resulting average duration of the liveness
        interruption (cut off at $800$ slots horizon).
        Observe
        that the peak for node $0$ fits well to the median observed in Figure~\ref{fig:cdfs-0}.
        The curves are smooth and allow for easy and reliable localization of the optimal $T_{\mathrm{delay}}$.}
    \label{fig:optimal-Tdelay}
\end{figure}

To understand whether the network propagation delay distribution is sufficiently well-behaved
for an adversary to reproducibly broadcast messages so that they arrive at roughly half
of nodes by a fixed deadline, we replicated the gossip network of Ethereum 2 \cite{eth2-spec-p2p}
and measured the network propagation delay of test `ping' packets from a designated sender
to all nodes. The implementation in the Rust programming language
used \texttt{libp2p}'s \texttt{Gossipsub} protocol and implementation,
as is used in Ethereum 2 \cite{eth2-spec-p2p}.

The gossip network comprised
$750$ nodes, each on an AWS EC2 \texttt{m6g.medium} instance
(with $50$ instances each in all $15$ AWS regions that supported \texttt{m6g.medium} as of 21-April-2021).
Each node initiated a connection with ten randomly chosen peers.
The five nodes with lowest instance ID were designated as senders
and continuously broadcasted beacon messages with inter-transmission
times uniformly distributed between zero and five seconds
over a period of $20$ minutes,
logging the time when each message was broadcast.
All nodes logged the time when a message was first received.
The network propagation delay was determined for each message and each receiving node.
The respective CDFs, %
\ie, what fraction of nodes have received a given message by a certain delay, is plotted as an example for a sample of messages
from
the first designated sender (node $0$) in Figure~\ref{fig:cdfs-0}
(together with the average CDF of all messages originating at node $0$).
(CDFs for the other four designated senders are omitted for brevity here.
They show similar behavior, just slightly shifted in time.)
It is apparent from the CDFs that depending on the location
of the node (nodes $0$, $1$, $2$, $3$, $4$ happened to be located in
\texttt{us-east-2}, \texttt{ap-northeast-1}, \texttt{us-east-1}, \texttt{ap-northeast-1}, \texttt{ap-northeast-2}, respectively)
both geographically as well as within the peer-to-peer network topology,
the median of the average CDF varies, but
considering messages
originating at a fixed sender,
the fraction of validators reached by
the median of the average CDF
is fairly concentrated around $1/2$.
This suggests that the adversary can indeed determine
$T_{\mathrm{delay}}$ so that with little dispersion
honest validators get split in two halves.

We simulated the attack for $\beta=0.15, m=128$, using the network propagation delay samples as a model for
random network delay.\footnote{Source code:
\url{https://github.com/tse-group/gasper-gossip-attack}}
Assigning the simulated adversary to one of the five designated senders
for all of the attack,
whenever the adversary broadcasts a sway vote, the propagation delays
to the honest committee members of the given slot
are sampled (without replacement) from the delays
of one randomly drawn message
of that designated sender.

To determine the optimal $T_{\mathrm{delay}}$,
we performed grid search (with $5\,\mathrm{ms}$ step size)
and for each $T_{\mathrm{delay}}$ simulated ten attacks in opportune epochs
and recorded (see Figure~\ref{fig:optimal-Tdelay})
how long the adversary was able to stall liveness
(terminating at a horizon of $800$ slots corresponding to $160$ minutes).
It is apparent that for the adversary in the position of each of the five designated senders
of the measurement experiment, different $T_{\mathrm{delay}}$ are optimal.
The optimal $T_{\mathrm{delay}}$ correspond well with the median of the average CDF (\cf Figure~\ref{fig:cdfs-0}).
As the curves are smooth and have a single distinct peak of
width $\approx5\,\mathrm{ms}$, the adversary can locate
the optimal $T_{\mathrm{delay}}$ well.
In particular, even with $T_{\mathrm{delay}}$ approximating the optimal value only
up to $10\,\mathrm{ms}$, the adversary can stall liveness for dozens of slots.
Recall that none of the adversarial actions are slashable protocol violations,
so the adversary can refine $T_{\mathrm{delay}}$ iteratively and launch this attack over and over.

\section{Reorg Attack Using Probabilistic Network Delay}
\label{sec:doubly-refined-reorg-attack}
\subsection{Motivation}

In Section~\ref{sec:refined-reorg-attack} we describe how an adversary might execute a $1$-reorg with only a single adversarial committee member's vote. In Section~\ref{sec:refined-liveness_attack} we show how an adversary can stall consensus and thus delay finality without adversarial control over network delay. By combining ideas from both attacks, we now describe an attack in which the adversary can execute a long-range reorg with vanishingly small stake and without control over network delay.

On a high level, the adversary avoids competing directly with honest validators of $(k-1)$ committees, as done in the reorg attack described in Section~\ref{sec:refined-reorg-attack}. Instead, the adversary uses the technique of Section~\ref{sec:refined-liveness_attack} to keep honest committee members split roughly in half by ensuring they have different views on what the current head of the chain is. This way, honest nodes work against each other and maintain a tie which the adversary can tip to their liking at any point using only a few votes.

\subsection{Refined Strategy Using Probabilistic Network Delay}

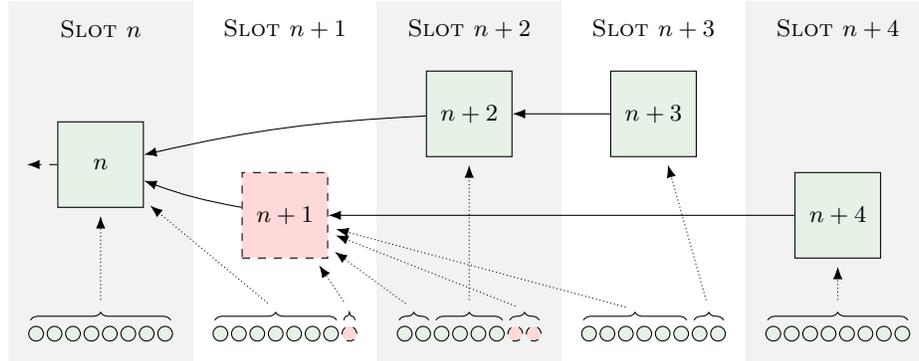
\begin{figure}[t]
    \centering
    \begin{tikzpicture}[
        _block/.style = {
            draw,
            minimum width=3.5em,
            minimum height=3.5em,
        },
        honestblock/.style = {
            _block,
            fill=green!50!black!10,
        },
        adversarialblock/.style = {
            _block,
            fill=red!15,
        },
        _vote/.style = {
            circle,
            draw,
            inner sep=0,
            minimum width=0.6em,
            minimum height=0.6em,
            font=\tiny,
        },
        honestvote/.style = {
            _vote,
            fill=green!50!black!10,
        },
        adversarialvote/.style = {
            _vote,
            fill=red!15,
        },
        hidden/.style = {
            dashed,
        },
        sloteven/.style = {
            fill=black!5,
            draw=none,
        },
        arrowvotes/.style = {
            -Latex,
            shorten <=3pt,
            shorten >=3pt,
            densely dotted,
        },
        groupvotes/.style = {
            decorate,
            decoration={
                brace,
                amplitude=3pt,
                raise=3pt,
            },
        },
        groupvotestip/.style = {
            midway,
            inner sep=0,
            yshift=7pt,
        },
        x=2.45cm,
        y=0.75cm,
    ]
        \footnotesize
    
        \draw [sloteven] (-0.5,2.9) rectangle (0.5,-3.5);
        \node at (0,2.4) {\textsc{Slot $n$}};
        
        \node at (1,2.4) {\textsc{Slot $n+1$}};
        
        \draw [sloteven] (1.5,2.9) rectangle (2.5,-3.5);
        \node at (2,2.4) {\textsc{Slot $n+2$}};
        
        \node at (3,2.4) {\textsc{Slot $n+3$}};
        
        \draw [sloteven] (3.5,2.9) rectangle (4.5,-3.5);
        \node at (4,2.4) {\textsc{Slot $n+4$}};

        \node [honestblock] (b1) at (0,0) {$n$};
        \node [adversarialblock,hidden] (b2) at (1,-0.9) {$n+1$};
        \node [honestblock] (b3) at (2,0.9) {$n+2$};
        \node [honestblock] (b4) at (3,0.9) {$n+3$};
        \node [honestblock] (b5) at (4,-0.9) {$n+4$};
        
        \draw [-Latex,dashed] (b1) -- (-0.40,0);
        \draw [-Latex,bend left=5] (b2) to (b1);
        \draw [-Latex,bend right=5] (b3) to (b1);
        \draw [-Latex] (b4) -- (b3);
        \draw [-Latex] (b5) -- (b2);
        
        \node [honestvote] (v11) at ($(0,-3)+(-0.35,0)$) {};
        \node [honestvote] (v12) at ($(0,-3)+(-0.25,0)$) {};
        \node [honestvote] (v13) at ($(0,-3)+(-0.15,0)$) {};
        \node [honestvote] (v14) at ($(0,-3)+(-0.05,0)$) {};
        \node [honestvote] (v15) at ($(0,-3)+(+0.05,0)$) {};
        \node [honestvote] (v16) at ($(0,-3)+(+0.15,0)$) {};
        \node [honestvote] (v17) at ($(0,-3)+(+0.25,0)$) {};
        \node [honestvote] (v18) at ($(0,-3)+(+0.35,0)$) {};
        
        \draw [groupvotes]
            (v11.north west) -- (v18.north east) node [groupvotestip] (q11) {};
        \draw [arrowvotes] (q11) -- (b1);
        
        \node [honestvote] (v21) at ($(1,-3)+(-0.35,0)$) {};
        \node [honestvote] (v22) at ($(1,-3)+(-0.25,0)$) {};
        \node [honestvote] (v23) at ($(1,-3)+(-0.15,0)$) {};
        \node [honestvote] (v24) at ($(1,-3)+(-0.05,0)$) {};
        \node [honestvote] (v25) at ($(1,-3)+(+0.05,0)$) {};
        \node [honestvote] (v26) at ($(1,-3)+(+0.15,0)$) {};
        \node [honestvote] (v27) at ($(1,-3)+(+0.25,0)$) {};
        \node [adversarialvote,hidden] (v28) at ($(1,-3)+(+0.35,0)$) {};
        
        \draw [groupvotes]
            (v21.north west) -- (v27.north east) node [groupvotestip] (q21) {};
        \draw [arrowvotes] (q21) -- (b1);
        \draw [groupvotes]
            (v28.north west) -- (v28.north east) node [groupvotestip] (q22) {};
        \draw [arrowvotes] (q22) -- (b2);
        
        \node [honestvote] (v31) at ($(2,-3)+(-0.35,0)$) {};
        \node [honestvote] (v32) at ($(2,-3)+(-0.25,0)$) {};
        \node [honestvote] (v33) at ($(2,-3)+(-0.15,0)$) {};
        \node [honestvote] (v34) at ($(2,-3)+(-0.05,0)$) {};
        \node [honestvote] (v35) at ($(2,-3)+(+0.05,0)$) {};
        \node [honestvote] (v36) at ($(2,-3)+(+0.15,0)$) {};
        \node [adversarialvote,hidden] (v37) at ($(2,-3)+(+0.25,0)$) {};
        \node [adversarialvote,hidden] (v38) at ($(2,-3)+(+0.35,0)$) {};
        
        \draw [groupvotes]
            (v31.north west) -- (v32.north east) node [groupvotestip] (q31) {};
        \draw [arrowvotes] (q31) -- (b2);
        \draw [groupvotes]
            (v33.north west) -- (v36.north east) node [groupvotestip] (q32) {};
        \draw [arrowvotes] (q32) -- (b3);
        \draw [groupvotes]
            (v37.north west) -- (v38.north east) node [groupvotestip] (q33) {};
        \draw [arrowvotes] (q33) -- (b2);
        
        \node [honestvote] (v41) at ($(3,-3)+(-0.35,0)$) {};
        \node [honestvote] (v42) at ($(3,-3)+(-0.25,0)$) {};
        \node [honestvote] (v43) at ($(3,-3)+(-0.15,0)$) {};
        \node [honestvote] (v44) at ($(3,-3)+(-0.05,0)$) {};
        \node [honestvote] (v45) at ($(3,-3)+(+0.05,0)$) {};
        \node [honestvote] (v46) at ($(3,-3)+(+0.15,0)$) {};
        \node [honestvote] (v47) at ($(3,-3)+(+0.25,0)$) {};
        \node [honestvote] (v48) at ($(3,-3)+(+0.35,0)$) {};
        
        \draw [groupvotes]
            (v41.north west) -- (v46.north east) node [groupvotestip] (q41) {};
        \draw [arrowvotes] (q41) -- (b2);
        \draw [groupvotes]
            (v47.north west) -- (v48.north east) node [groupvotestip] (q42) {};
        \draw [arrowvotes] (q42) -- (b4);
        
        \node [honestvote] (v51) at ($(4,-3)+(-0.35,0)$) {};
        \node [honestvote] (v52) at ($(4,-3)+(-0.25,0)$) {};
        \node [honestvote] (v53) at ($(4,-3)+(-0.15,0)$) {};
        \node [honestvote] (v54) at ($(4,-3)+(-0.05,0)$) {};
        \node [honestvote] (v55) at ($(4,-3)+(+0.05,0)$) {};
        \node [honestvote] (v56) at ($(4,-3)+(+0.15,0)$) {};
        \node [honestvote] (v57) at ($(4,-3)+(+0.25,0)$) {};
        \node [honestvote] (v58) at ($(4,-3)+(+0.35,0)$) {};
        
        \draw [groupvotes]
            (v51.north west) -- (v58.north east) node [groupvotestip] (q51) {};
        \draw [arrowvotes] (q51) -- (b5);

    \end{tikzpicture}
    \caption{Example of a $2$-reorg combining refined reorgs and balancing strategies:
    In slot $n+1$
    the adversary privately creates block $n+1$ on block $n$ and withholds adversarial votes on it. Honest validators of slot $n+1$ 
    attest to block $n$.
    In slot $n+2$,
    an honest proposer 
    builds
    block $n+2$
    on block $n$.
    The adversary releases block $n+1$ and one of the withheld votes in such a way that roughly half of honest committee members vote for blocks $n+1$ and $n+2$, respectively.
    If the adversary has tight control over network delays, they can effect that
    block $n+2$ 
    has one more vote than block $n+1$.
    Without adversarial control of delays, 
    a vanishing fraction of
    adversarial votes still suffices to rebalance accordingly.
    In slot $n+3$, the honest proposer views block $n+2$ leading and proposes block $n+3$ off it.
    The adversary releases 
    two votes 
    voting for block $n+1$ in such a way that a majority of honest committee members 
    vote for block $n+1$, breaking the tie and completing the $2$-reorg which orphaned blocks $n+2$ and $n+3$ in slot $n+4$.}
    \label{fig:exante-reorg-refined}
\end{figure}

Consider Figure~\ref{fig:exante-reorg-refined}, in which the adversary is the proposer of slot $n+1$.
We describe the strategy where the adversary executes a $2$-reorg and analyze how many validators the adversary needs to control, depending on our assumption on the adversary's control over the network:
\begin{enumerate}
    \item First, in slot $n+1$ the adversary privately builds block $n+1$ on top of the current head of the chain, block $n$. Further, the adversary privately votes for block $n+1$ using an attestation from slot $n+1$.

    \item In the next slot, the proposer of block $n+2$ builds on block $n$ because they have not seen block $n+1$.
    Before honest validators in slot $n+2$ attest, the adversary releases block $n+1$, along with the withheld attestation,
    in such a way that
    roughly half of honest committee members of slot $n+2$ attest before they see the sway vote (and thus vote for block $n+2$ as the current head), and the other half
    sees block $n+1$ as leading due to the attestation from slot $n+1$ and thus votes
    for block $n+1$ as the current head.

    If the adversary has control over the network delay, as assumed in \cite{ebbandflow,ethresearch-balancing-attack}, then it can target the release of the withheld block and vote such that block $n+2$ accumulates exactly one more attestation than block $n+1$.
    If network delay is instead probabilistic, as in Section~\ref{sec:refined-liveness_attack},
    then the adversary needs to spend $O(\sqrt{W_{\mathrm{honest}}})$ adversarial votes to rebalance the gap in votes.
    
    In the case of a $k$-reorg,
    this step is repeated for the first $(k-1)$ slots.
    
    \item Since slot $n+3$ is the last slot of the reorg attack, we use the insight of Section~\ref{sec:refined-reorg-attack} that the adversary does not have to wait for honest votes to take place and rebalance them, but instead can sway validators towards the adversarial chain as soon as the honest proposal for this slot was created.
    So, in slot $n+3$, the current proposer views block $n+2$ as leading and thus builds block $n+3$ on it. Finally, the adversary releases two withheld attestations such that a majority of honest committee members of slot $n+3$ views them before attesting. Thus, a majority of validators votes for block $n+1$ as head of the chain.
    Remember that the fork choice rule only considers attestations at least one slot old.
    
    \item Lastly, in slot $n+4$ the proposer views block $n+1$ as leading and thus builds block $n+4$ on block $n+1$. This completes the $2$-reorg and orphans blocks $n+2 $ and $n+3$. 
    
\end{enumerate}

For $1$-reorg the adversary needs to control a single validator in the same slot they propose their block. For reorg lengths $k>1$, the number of adversarial validators required depends on the level of control over network delays.
If delays are under adversarial control,
then $(2k-1)$ adversarial validators suffice
for a $k$-reorg, an amount linear in the reorg length only, but independent of
the size of the validator set.
If instead network delay is probabilistic rather than under adversarial control,
a vanishingly small fraction $O(1/\sqrt{W_{\mathrm{honest}}})$ of adversarial validators
suffices to perform the necessary rebalancing to maintain the tie
throughout the first $(k-1)$ slots of the $k$-reorg,
leading to an overall requirement of
$O(k \sqrt{W_{\mathrm{honest}}})$ adversarial votes.
Thus, large
stakers can easily execute long-range reorg attacks. To illustrate the severe reduction of attacking conditions, consider the following: Under adversarial network delay,
an adversary can perform a $10$-reorg by merely controlling $19$ validators.

\section{Discussion}
\label{sec:discussion}
\subsection{Ex Ante Vs Ex Post Reorgs}
\label{sec:discussion-ex-ante-ex-post}

Typically reorgs refer to an attack in which the adversary observes a block that they subsequently attempt to fork out. We call this an \emph{ex post} reorg attack. The reorg attacks we describe are different in nature. Here, the adversary attempts to fork out a future block that is unknown to the adversary at the start of the attack. We call this an \emph{ex ante} reorg attack.

In an ex post reorg attack, the adversary typically targets a block 
with
abnormally large rewards that the adversary 
seeks to capture for themselves. In the context of Bitcoin it could be a block that contains transactions paying extraordinary amounts of 
fees, also referred to as `whale transactions' \cite{whaletx}. In the context of Ethereum it could be blocks containing
large MEV opportunities. Upon observing a lucrative block, the adversary attempts to capture it retrospectively. In PoS Ethereum this proves to be exceptionally difficult for non-majority actors due to the fact that the block the adversary wishes to orphan quickly accrues attestations from committee members in parallel. Each attestation adds weight to the block in question, which in turn is considered by the fork-choice rule LMD GHOST to determine the head of the chain. In short, no technique is known for non-majority adversaries to perform ex post reorg attacks reliably.

In contrast, ex ante reorg attacks are currently very much possible in PoS Ethereum, as this paper shows. The adversary overcomes the `power of many parallel attestations' by exploiting LMD GHOST as described in Sections~\ref{sec:refined-reorg-attack} and~\ref{sec:doubly-refined-reorg-attack}. Intuitively, this is enabled by 
tricking honest validators into contrary views of the chain such that a handful of adversary validators are sufficient to tip the chain to their favor and thus successfully perform reorgs of sizable length. As a consequence of the different nature of the attack, the adversary's motivations to attack are different. In an ex ante reorg the adversary cannot observe valuable blocks and orphan them ex post, but must find other strategies to extract more value from it than it could from making an honest proposal, one of which is discussed in Section~\ref{sec:discussion-mev-sub}.

\subsection{Reaping Higher Fees and MEV Via The Attack}
\label{sec:discussion-mev-sub}

Maximal Extractable Value (MEV, formerly Miner Extractable Value \cite{flashboys}) represents a third source of profits for block producers, along with the proposer and attester rewards as well as transaction fees. MEV in PoS Ethereum captures the block proposer's action space to extract value by strategically including and ordering transactions in a given block. Common MEV opportunities include arbitraging a trade, frontrunning it to earn greater profits, or tailing liquidation events to buy the collateralized assets backing the defaulting position.

MEV opportunities grow with an increasing amount of pending transactions since more possible transaction order combinations exist. At the same time, the adversary is able to choose from a larger set of pending transactions those earning them the highest fees. More time between blocks then implies weakly more extractable MEV and transaction fees, which in turn implies more profits for the block proposer. The reorg attacks described in this paper can be interpreted as buying the adversary more time to construct their block.

With $k$-reorgs, it is possible for the malicious proposer to extend their listening period to up to $12k$ seconds (refined reorg strategy from Section~\ref{sec:refined-reorg-attack}), the 12 seconds elapsed between the previous block produced and their own slot, as well as $12(k-1)$ more seconds until the next honest block is included in the canonical chain. (The 2$\Delta$ duration introduced in Section~\ref{sec:prelims-protocol} is set to 12 seconds in the PoS Ethereum implementation.) With $k$-reorgs in less idealized scenarios, as described in Section \ref{sec:refined-reorg-attack}, the adversary only gains an additional 12 seconds of listening time (24 seconds in total). This is due to the fact that in the refined strategy using probabilistic network delay the adversary always releases the private block early (irrespective of reorg length $k$) to split honest committees roughly in half.

Further, the adversary may listen to honest blocks they wish to orpahn, and capture their MEV should they find better opportunities than the adversary themselves. Interestingly, the adversary may also simply release their block late, without attempting a reorg, to increase their listening time and ultimately rewards.

\subsection{Reorgs Cause Attestation Overflow}
\label{sec:discussion-reorgs-attestation-overflow}

While reorg attacks weakly benefit those who launch them,
consensus degradation may be obtained as an unintended side-effect of the reorg.

Validators in a slot committee are distributed among a number of subcommittees. With a target subcommittee size of $128$ and 
currently
$230{,}000$ active validators, 
$\approx 57$ subcommittees are formed per slot. In the current implementation of PoS Ethereum, all identical votes from the same subcommittee may be aggregated into one `summary' vote, lightening the block size. 
A block may include up to $128$ such aggregates. 
Ideally,
with all validators voting correctly and on time, the next block need only feature $57$ aggregates, one per subcommittee. In practice, we observe such a number of large aggregates (summarizing many votes) in the block, with most validators voting identically, along with some aggregates summarizing other votes from validators who may have suffered from latency issues and voted identically, albeit wrongly. Suboptimal packing of the aggregates or adversarial voting behavior may also contribute to filling up the available slots for aggregates in the block.
In the case of a reorg, deconfirmed aggregates return into the mempool and need to be included
in future blocks. Even for short-range reorgs this can lead to congestion in the sense
that many more aggregates wait to be included than there is space available in blocks.

Votes state their view of the current target of the FFG mechanism. 
A target vote is valid only if it is
included in a block no later than $32$ slots after the attesting slot. By reorging blocks, an attacker strains the capacity of the chain to include these valid votes. In the worst case, finalization is fully delayed whenever more than $1/3 - \beta$ of valid honest votes do not manage to be included.

\subsection{Delaying Finality}
\label{sec:discussion-delaying-finality-sub}

Our attacks also enable \emph{a priori} malign actors,
perhaps ideologically motivated,
to delay and in some cases outright stall consensus decisions.
The refined attack of Section~\ref{sec:appendix-attack-gasper-high-level}
gives the adversary a tool to do just that,
even if the adversary cannot control message propagation delays
(which instead are assumed to be probabilistic).
Furthermore, in the regime of many validators,
a vanishing fraction of adversarial stake suffices to mount the attack.
The attack of Section~\ref{sec:doubly-refined-reorg-attack} enables
long-range reorgs of the chain constituting consensus.
The consequences are two-fold.
Readily, transaction confirmation in the LMD GHOST part of the protocol
gets delayed. Transactions might enter/leave the LMD GHOST chain
multiple times before eventually settling.
This causes
uncertainty and delay
for users
who consider
a transaction confirmed once 
it has stabilized in the LMD GHOST chain.
Furthermore, the adversary can use reorgs, as proposed
in \cite{lowcostreorgs},
to destabilize epoch boundary blocks.
No epoch boundary block might then get the necessary number of FFG votes to
become justified, which delays finality by at least an epoch
and thus creates delay for users who rely on the finalized ledger.

\section*{Acknowledgment}
JN, ENT and DT are supported by a gift from the Ethereum Foundation.
JN is supported by the Reed-Hodgson Stanford Graduate Fellowship.
ENT is supported by the Stanford Center for Blockchain Research.

\bibliographystyle{splncs04}
\bibliography{references}

\appendix

\end{document}